\documentclass[11pt,a4paper]{article}

\usepackage{graphicx}
\usepackage{epstopdf}
\usepackage{amsmath}
\usepackage{url}
\usepackage{xr}
\usepackage{a4wide}
\usepackage{natbib}
\usepackage{soul}
\usepackage{color}
\usepackage{subcaption}
\usepackage{placeins}

\graphicspath{{figures//}}
\title{Graph-based data integration predicts long-range regulatory interactions across the human genome}

\author{Sofie Demeyer\,$^{1}$ and Tom Michoel\,$^{1,\ast}$}

\begin{document}

\maketitle

$^1$ Division of Genetics and Genomics, The Roslin Institute, The University of Edinburgh, Midlothian EH25 9RG, Scotland, United Kingdom

\medskip

$^\ast$Corresponding author: tom.michoel@roslin.ed.ac.uk

\medskip

\textbf{Running title:} Graph-based prediction of long-range regulatory interactions

\newpage

\begin{abstract}
  Transcriptional regulation of gene expression is one of the main processes that affect cell diversification from a single set of genes. Regulatory proteins often interact with DNA regions located distally from the transcription start sites (TSS) of the genes. We developed a computational method that combines open chromatin and gene expression information for a large number of cell types to identify these distal regulatory elements. Our method builds correlation graphs for publicly available DNase-seq and exon array datasets with matching samples and uses graph-based methods to filter findings supported by multiple datasets and remove indirect interactions.  The resulting set of interactions was validated with both anecdotal information of known long-range interactions and unbiased experimental data deduced from Hi-C and CAGE experiments. Our results provide a novel set of high-confidence candidate open chromatin regions involved in gene regulation, often located several Mb away from the TSS of their target gene.\\

\end{abstract}

\section{Introduction}

The central dogma of molecular biology states that genetic information flows from DNA to RNA to proteins. The rate and level at which the different genes are transcribed determine the functionality of each cell. This process is controlled by regulatory proteins binding or unbinding to/from specific DNA regions, in response to signals coming from within a cell, from neighboring cells or directly from the external environment.
While some of the proteins act locally, i.e. close to the transcription start site (TSS) of the genes, others are known to bind to distal regions, even across gene boundaries (\citet{akalin2009, noonan2010, ernst2011}). Systematically identifying these distal regulatory elements and their target genes remains one of the principal challenges of regulatory genomics.

With the help of the ENCODE consortium (\citet{encode2011,thurman2012}), large amounts of cellular data have been made publicly available. This includes both gene expression data, such as exon arrays (\citet{kapur2007}) and RNA sequencing data (\citet{Nagalakshmi2008}),  and open chromatin data, more specifically DNase-seq data. DNase sequencing (\citet{boyle2008}) is a genome-wide extension of the DNase I footprinting method (\citet{galas1978}) and identifies open chromatin regions that are sensitive to cleavage by the DNase I enzyme and thus accessible to DNA-binding proteins (\citet{wu1980}).

Recently, linking open chromatin and gene expression information has gained attention; see \citet{xi2007, natarajan2012, sheffield2013, marstrand2014}. Each of these studies presents a computational method that combines DNase-seq data and gene expression data for different cell types to identify correlations between the two. However, all of these studies take only a limited number of cell types into account and are limited to open chromatin regions located in the `proximity' of the genes (ranging between 100kb and 500kb). Nevertheless, there are known chromatin interactions between sites located several Mb away from each other (\citet{li2012}).

We propose a computational method to identify interactions between open chromatin regions and genes by combining open chromatin and gene expression information for more than 100 cell types. We do not limit the area of interest around the genes, but take into account all open chromatin regions of the whole chromosome. In this way we aim at discovering novel regulatory mechanisms, including unknown open chromatin regions interacting with genes located several Mb away.

Essentially, the proposed method obtains more accurate results by combining multiple datasets. After calculating the correlations for each dataset separately, the results are combined in a graph-based manner. This offers a distinct advantage in that the actual open chromatin and gene expression values need not to be compared directly across datasets, eliminating the need to jointly normalize the different input datasets. In addition, a similar graph-based method, used to identify open chromatin regions interacting with multiple genes, eliminates the indirect interactions.

The predicted interactions were validated with both anecdotal information of known interactions and unbiased validation sets (more specifically Hi-C (\citet{jin2013}) and CAGE data (\citet{andersson2014})).

\section{Results}

\subsection{A graph-based data integration methodology}

For 103 cell types, we collected DNase-seq peak data and gene expression data from the human ENCODE database (\citet{encode2011}). DNase-seq data was collected in 100bp bins which we  refer to as DNase hypersensitive sites (``DHS'') (see Methods). Gene expression levels were available in the form of exon array data, which came from two different sources: 66 cell lines were collected by the University of Washington and reported exon-level data, and 37 cell lines were combined data collected by both the University of Washington and Duke University and reported gene-level data. Instead of attempting to normalise all samples to a common scale, both sets were treated as separate datasets (henceforth called the ``UW'' and ``DukeUW'' datasets) and graph-based methods were used to integrate them.

\begin{figure}[ht!]
  \centering
  \includegraphics[width=\linewidth]{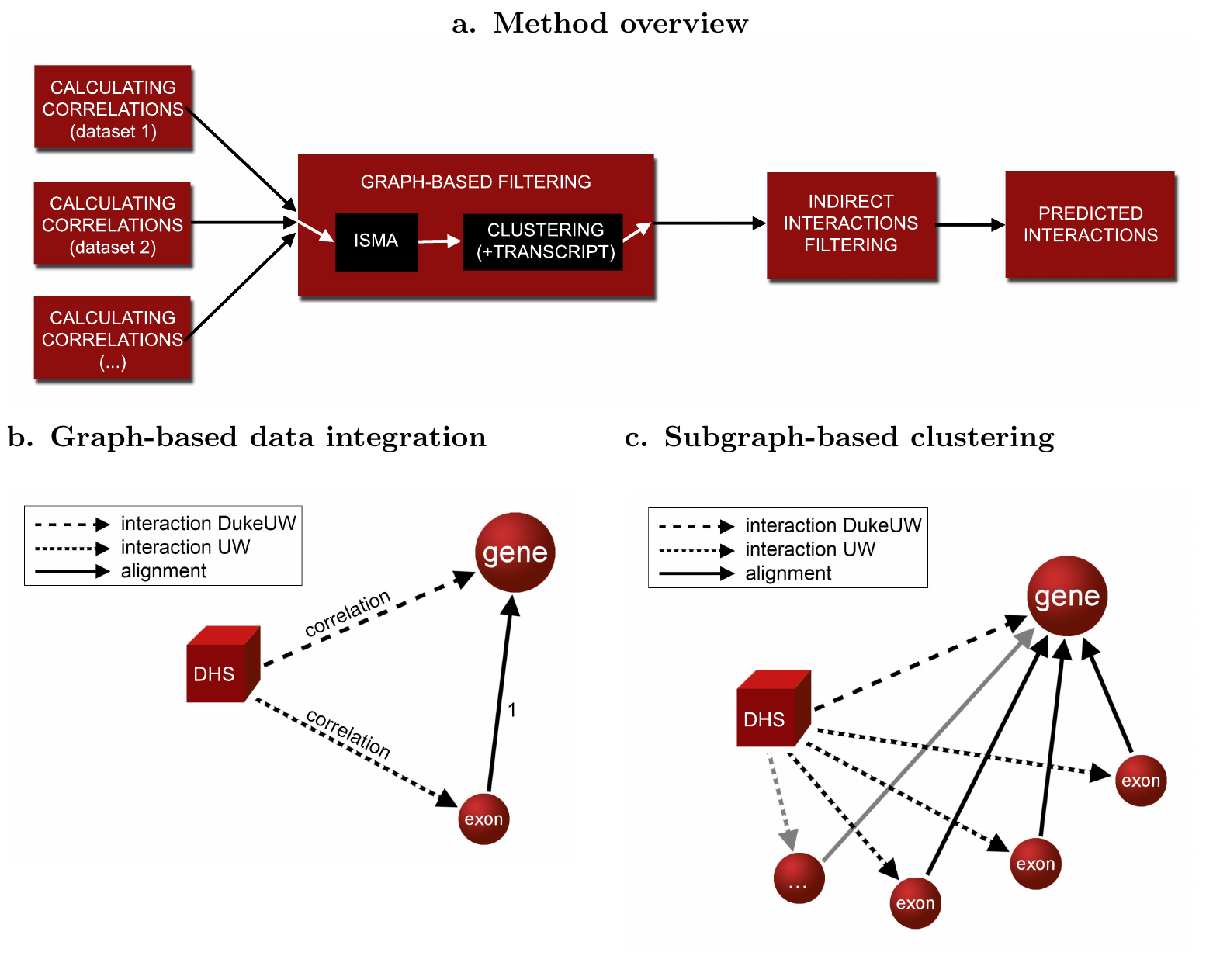}
  \caption{\textbf{Computational method overview.} \textbf{a.} DHS--gene regulatory interactions are predicted by a computational pipeline that combines significant correlation values calculated separately for multiple datasets (left) using graph-based data integration (middle) and indirect interaction filtering methods (right). \textbf{b.} The graph-based data integration method initially uses subgraph pattern matching to identify consistent predictions between datasets, in this case consistent significant DHS--gene correlations in the DukeUW dataset and significant DHS--exon correlations in the UW dataset. \textbf{c.} In a second step, subgraph clustering is used to group related subgraphs, in this case to detect DHS--gene correlations in the DukeUW dataset consistent with groups of DHS--exon correlations in the UW dataset, where all exons combine to form known transcripts for that gene.}
    \label{fig:overview_method}
\end{figure}

A schematic overview of the method is depicted in Figure \ref{fig:overview_method}. First, for each dataset, absolute Spearman correlations were calculated between all pairs of DHSs and exons/genes lying on the same chromosome. To identify the most significant interactions, an empirical null distribution was calculated from randomly permuted data and all correlations with empirical FDR values below 10\% were retained (see Methods for details). This resulted in a set of DHS--gene interactions for each dataset, together with their correlation values which serve as an interaction weight or quality score. Next, a weighted edge-colored network was constructed with all DHSs, genes and exons as nodes and three types of edges: interactions from the DukeUW dataset, interactions from the UW dataset and alignment links between the different datasets (in this case mapping exons to co-located genes). This network was analyzed with the ISMA algorithm, a highly efficient subgraph matching algorithm (\citet{demeyer2013}), in order to identify all subgraphs that represent an interaction occurring in both datasets (Figure \ref{fig:overview_method}b). For each subgraph instance, a quality score was calculated as the geometric mean of its edge weights. Next, to reconstruct unique DHS--gene interactions, we identified subgraph clusters (\citet{michoel2012}) (Figure \ref{fig:overview_method}c) and retained only those clusters (i.e.\ DHS--gene interactions) for which the set of exons matches with known gene transcripts (see Methods for details). The final quality score is calculated as the maximum quality score of the 3-node subgraphs in a cluster.

Table \ref{tab:number_interactions} shows the number of interactions, genes/exons and DHSs found after each calculation step described above  (see also Supplementary Table 2 for a complete list of the number of interactions per chromosome). As expected the graph-based filtering considerably reduces the number of predicted interactions. Hereafter, we will demonstrate that this filtering also improves the quality of the predictions.

\begin{table}
\begin{center}
\begin{tabular}[ht]{|l|r|r|r|}
\hline
	& \# `interactions' & \# genes & \# DHSs\\
\hline
original interactions (DukeUW)	& 7 129 048 & 13 757 & 759 123\\
original interactions (UW)	& 890 706 846 & 271 780* & 1 644 914\\
after ISMA filtering	& 21 234 602 & 6 953 & 247 495\\
&  & 72 652* &\\
after clustering	& 2 167 676 & 6 953 & 247 495\\
after transcript filtering	& 505 716 & 1 833 & 126 543\\
\hline
\end{tabular}
\end{center}
\caption{\textbf{Number of `interactions' after each calculation step.} `Interactions' represent either actual interactions between a DHS and a gene/exon (original interactions), 3-node motifs (after ISMA filtering) or single interactions between a DHS and a gene as clusters (after clustering and after transcript filtering). The genes represent either actual genes or exons (marked with a *).}
\label{tab:number_interactions}
\end{table}

\subsection{Filtering indirect interactions}

\begin{figure}[ht!]
    \centering
    \includegraphics[width=\linewidth]{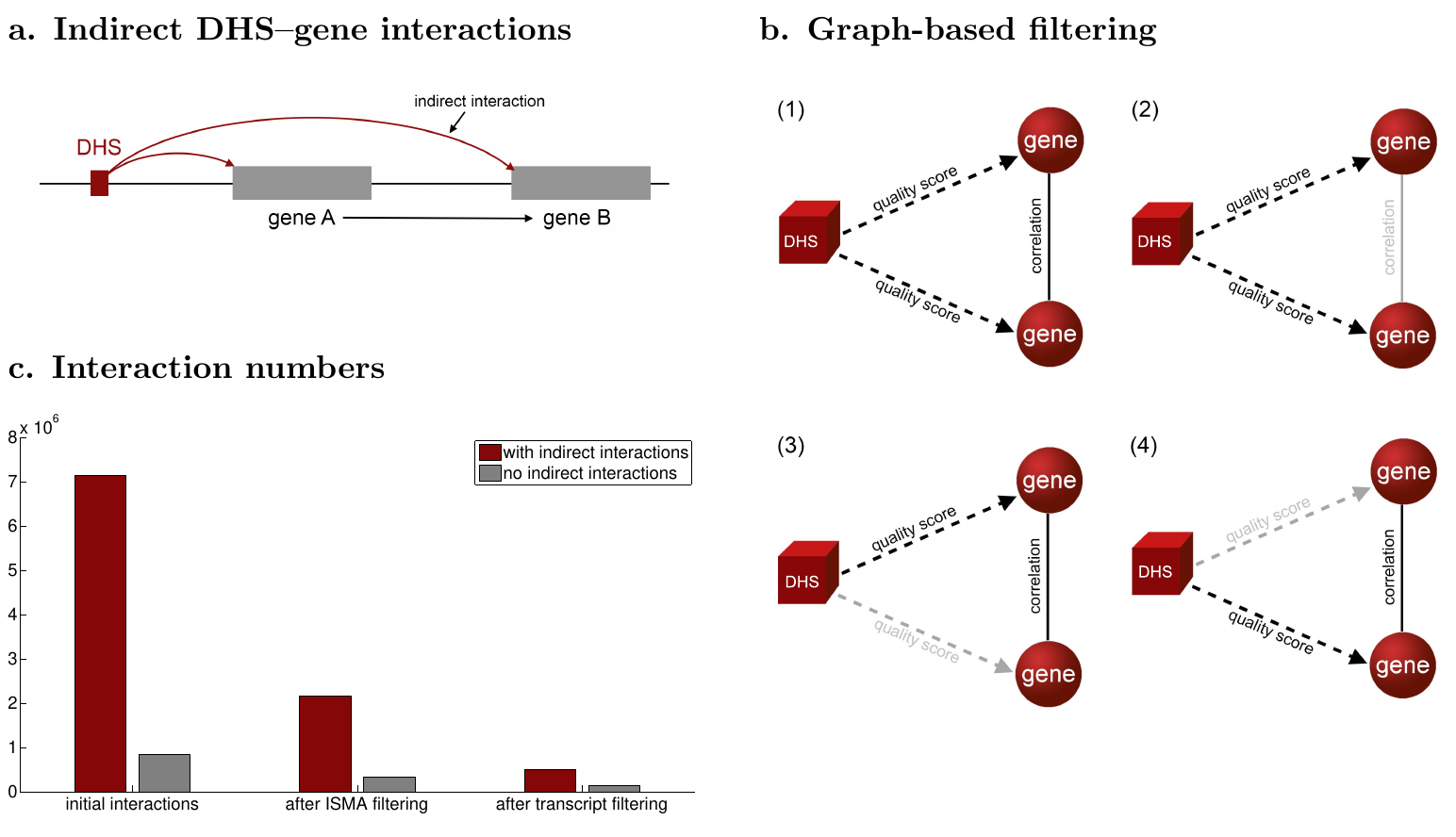}
    \caption{\textbf{Filtering indirect interactions.} \textbf{a.} An indirect interaction between a DHS and gene B will be inferred if gene B is regulated by gene A which in turn is regulated by the DHS. \textbf{b.} Subgraph pattern matching is used to find pairs of significant DHS--gene correlations where both genes are significantly correlated as well (1); to filter indirect interactions we remove from such subgraphs the most weakly supported edge (2--4). \textbf{c.} For each step of the pipeline the number of interactions is shown before and after indirect interaction filtering.}
    \label{fig:indirect_interactions}
\end{figure}

We calculated DHS--gene interactions from significant correlations between open chromatin areas and genes. However, some of these interactions might be of an indirect nature. Suppose, for example, a DHS is a \textit{bona fide} regulatory element for gene A, but this gene A in turn regulates the expression level of another gene B. While correlation data alone might suggest that this DHS is also a regulatory element for gene B, it is actually a consequence of both genes interacting (Figure \ref{fig:indirect_interactions}a). Our graph-based method to eliminate indirect DHS-gene correlations begins with the construction of another edge-colored network, this time containing DHS-gene interactions inferred from the previous analysis and gene-gene correlation interactions (see Methods for details). Then, we again used the ISMA algorithm (\citet{demeyer2013}) to identify subgraphs that represent a single DHS interacting with two mutually correlated genes (Figure \ref{fig:indirect_interactions}b(1)). Finally, we removed from these subgraphs the most weakly supported edge (Figure \ref{fig:indirect_interactions}b(2--4)) and reassembled the DHS--gene interaction network from the remaining interactions, i.e.\ a DHS is predicted as a regulatory element for two different genes if and only if the strength of correlation between the DHS and both genes is greater than the mutual correlation between the genes.

A special case of an indirect interaction occurs when a DHS is located inside a gene body. Because expressed genes are located in open chromatin regions (\citet{natarajan2012}) a large number of this type of interactions are predicted, while only some of them correspond to true regulatory interactions. We opted to remove all (looping) interactions between DHSs and co-located genes from the results. Notice that genuine enhancers that have been found within gene bodies (\cite{arnold2013}) will still be predicted if the quality score of this interaction is higher than the weight of one of the two other links in the motif.

Figure \ref{fig:indirect_interactions}c shows for each step of the computational pipeline the number of interactions before and after indirect interaction filtering (see Supplementary Table \ref{stab:number_indirect_interactions} for the numbers per chromosome). In each step of the calculations, a large number of indirect interactions are filtered out. Some of these interactions were already identified by previous filtering as the proportion of the number of interactions without the indirect ones to the total number of interactions increases with additional filtering (see Supplementary Table \ref{stab:number_interactions}). Only 11.99\% of the initial interactions remain after indirect filtering. After ISMA filtering this percentage increases to 15.76\%, and after transcript filtering to 27.17\%.

\subsection{Extensive long-range interactions remain after filtering}

Previous joint analyses of DNase I hypersensitivity and gene expression data have always focused on open chromatin regions located in the proximity of transcription start sites (TSSs) of genes using arbitrary distance cut-offs (varying between studies from 100--500kb) (\citet{degner2012, natarajan2012, sheffield2013, marstrand2014}). Here we predicted interactions across whole chromosomes and asked whether the data supports the use of hard cut-offs proximal to the TSS.

\begin{figure}[ht!]
    \centering
    
    \includegraphics[width=\linewidth]{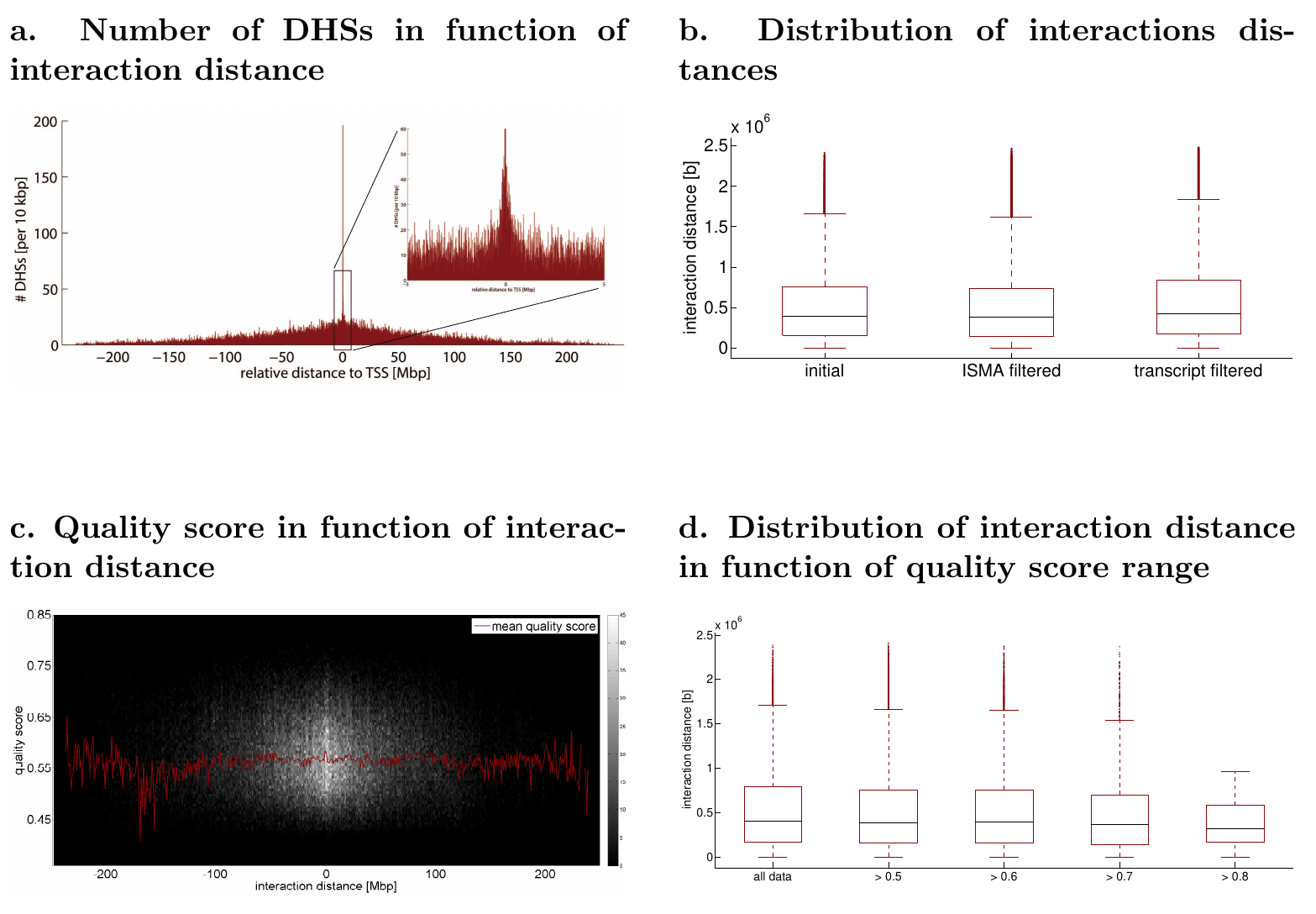}
		
    \caption{\textbf{Long-range interactions.} \textbf{a.} Number of interacting DHSs in function of the relative distance to the TSS. While there are more interacting DHSs close to the TSSs of the genes, still interactions are found between DHSs and genes located several Mb away. \textbf{b.} The distribution of the interaction distance does not change in the different filtering steps. This was also validated by the Wilcoxon rank test. \textbf{c.} The mean of quality scores (red line) remains more or less the same across the interaction distances. Again, a higher concentration of interactions was observed around the TSS. \textbf{d.} The distribution of the interaction distance does not change remarkably for different quality score cut-off values. This was confirmed with the Wilcoxon rank test.}
    \label{fig:DHS_ifo_TSS}
\end{figure}

Figure \ref{fig:DHS_ifo_TSS}a shows the number of interactions in function of the distance between the open chromatin areas and TSS of their interacting genes, derived from the list of interactions after all filtering steps (i.e.\ ISMA, transcript and indirect interactions). Although the number of interactions decreases with increasing distance, a high number of interactions occur between sites located several Mb away from each other (see also Supplementary Table \ref{stab:long_range}). Similar results (see Supplementary Figure \ref{sfig:DHS_ifo_TSS}) were obtained for the intermediate lists of interactions, i.e.\ before or between the filtering steps, and no significant difference in the distributions of interaction distances was observed between the interaction lists (Figure \ref{fig:DHS_ifo_TSS}b).

Next, it was investigated how the interaction quality scores are related to the distances between the DHSs and the genes. As Figure \ref{fig:DHS_ifo_TSS}c shows, the average quality score does not depend strongly on distance and interactions with high quality scores are found at all distances. Furthermore, there are no significant differences in the distance distribution at various score cut-offs, except at the very stringent threshold ($>$0.8) (Figure \ref{fig:DHS_ifo_TSS}d). Again, similar results are obtained for the intermediate lists of interactions (see Supplementary Figure \ref{sfig:QS_box} and \ref{sfig:QS_ifo_TSS}).

\subsection{Chromosome capture data systematically validates predicted interactions}

The predicted DHS--gene interactions were validated with chromosome conformation capture data. Chromosome Conformation Capture (3C) is an experimental technique to analyse the three-dimensional organization of chromosomes which reveals distal regions interacting with gene promoter regions (\citet{dekker2002, tolhuis2002, wei2011, dewit2012}). To increase the throughput of quantifying chromosomal interactions, a number of 3C-related techniques have been developed such as Circular 3C (4C) (\citet{zhao2006}), Carbon-Copy 3C (5C) (\citet{dostie2006, sanyal2012}) and Hi-C (\citet{lieberman2009}).  When a DHS is predicted to (functionally) interact with a certain gene by our method and it is found by chromosome capture data to lie in a chromosomal region that physically interacts with that gene's promoter, we considered this a validation of the predicted interaction.

\subsubsection{Genome-wide 3C (Hi-C) validation}
\label{sec:genome-wide-3c}

Genome-wide Hi-C data allows for the most systematic validation of predicted DHS--gene interactions. Here we used a Hi-C dataset (the `gold standard') from IMR90 (primary human fibroblast) cells which contained 57,059 interactions between 49,394 regions (1-20 kbp) centered on the \textit{cis}-elements annotated in the IMR90 cell genome and the promoter regions of 9181 genes with a FDR of 10\%, reporting only interactions within a 2 Mb distance (\citet{jin2013}). Following established protocols from the network inference field (\citet{stolovitzky2009lessons}), we only considered DHS--gene predictions within the gold standard space (i.e.\ the set of all possible interactions between DHSs and genes from the gold standard that are not located more than 2Mb from each other) and we counted a predicted interaction as a \emph{true positive} (TP) if it indeed appeared in the gold standard and as a \emph{false positive} if it did not (see Methods for details). We considered predictions at various quality score cut-offs and calculated precision (the proportion of TP in the predicted set) and recall or sensitivity (the proportion of the gold standard that was correctly predicted) at each cut-off value.

\begin{figure}[ht!]
	
  \includegraphics[width=\linewidth]{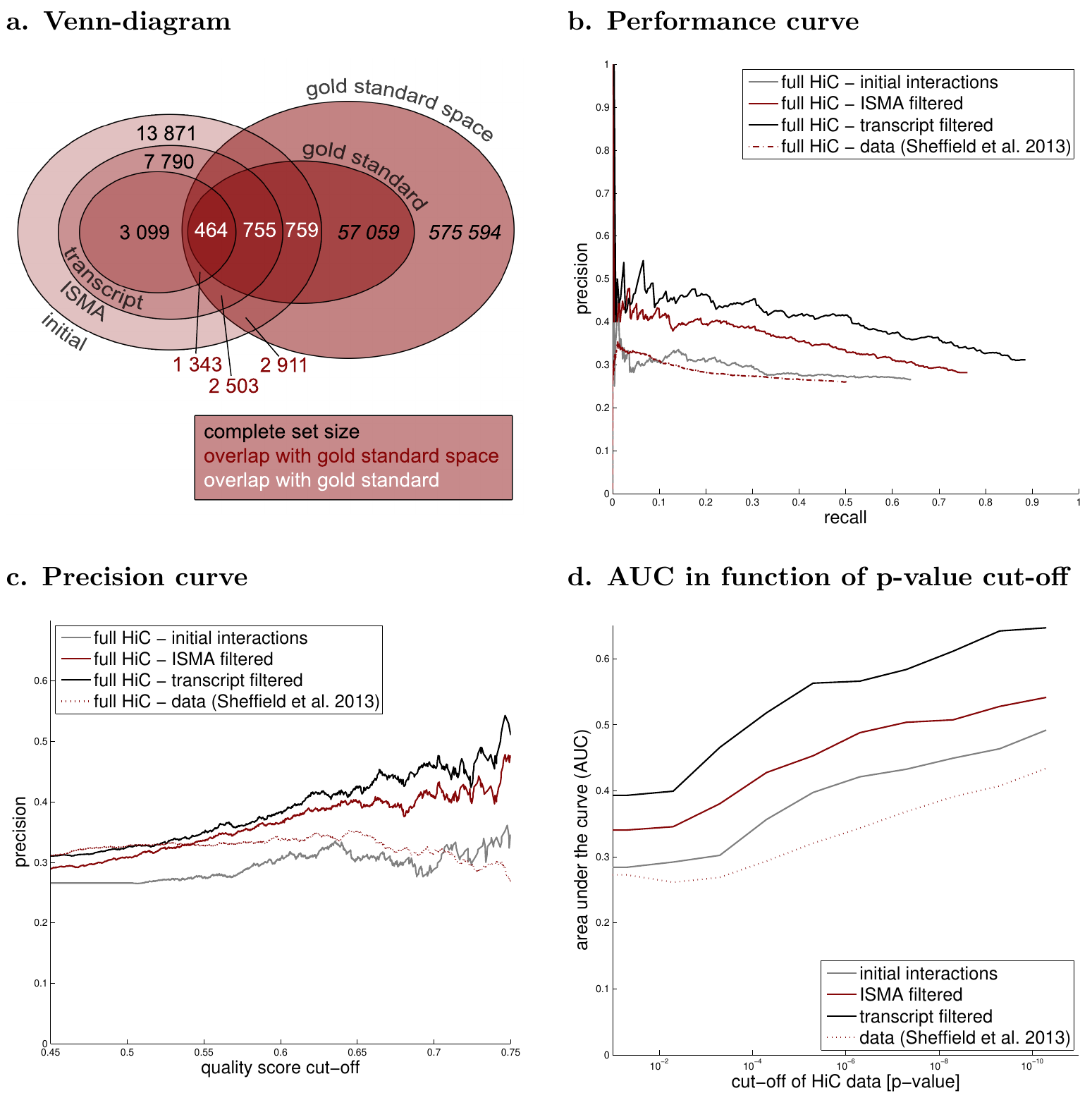}
	
  \caption{\textbf{Validation with Hi-C data.} The gold standard consists of all interactions found in the Hi-C experiments. \textbf{a.} To asses the meaningfulness of this validation, it was investigated whether a fair percentage of our predictions are present in the gold standard space. \textbf{b.} The performance curve shows the precision in function of the recall for different quality score cut-off values. \textbf{c.} The precision curve shows an increasing precision with increasing quality score cut-off values. \textbf{d.} The p-value cut-off for the Hi-C data of the gold standard was decreased, resulting in more stringent validation sets. The area under the performance curve (AUC) is shown in function of the p-value cut-off.}
  \label{fig:performance_HiC}
\end{figure}

Reliable estimation of true and false positive rates requires a large gold standard space (relative to the gold standard itself) and a sufficient overlap between the predictions and the gold standard (space). Figure \ref{fig:performance_HiC}a shows that the gold standard space is indeed ten-fold larger than the gold standard and that after
all filtering, 43.34\% of the predicted interactions, within the same 2Mb range as the gold standard, lie in the gold standard space, justifying our validation method. For the initial predictions and the predictions after ISMA filtering these percentages are respectively 20.99\% and 32.13\%.

Figure \ref{fig:performance_HiC}b shows the performance curves, i.e.\ the precision in function of the recall, of our predicted interactions at different stages of the prediction pipeline. For comparison, we also performed the same validation on the set of long-range interactions reported by \citet{sheffield2013}. As expected, the performance of our unfiltered list of interactions and the predictions of \citet{sheffield2013} are comparable since both use the Spearman correlation as a quality score on a comparable number of cell types, the only difference being the filtering of indirect interactions in our method and the limitation to distances less than 500kb by \citet{sheffield2013}. However a clear improvement in performance is seen for the filtered interaction lists which result in $\sim$1.5-fold increase in precision at the same level of recall. The same result is observed if we plot precision as a function of the quality score cut-off (Figure \ref{fig:performance_HiC}c).
 
Because the Hi-C data is itself subject to noise and potentially contains numerous false positive interactions, we used the confidence $p$-values reported by \citet{jin2013} to construct gold standards from increasingly stringent Hi-C interactions and repeated the same validation experiments. Using the area under the recall precision curve (AUC) as an overall performance measure (\citet{stolovitzky2009lessons}), we found that prediction quality indeed increased significantly for the more stringent sets and that the relative performance of the different filtered sets was conserved across the entire range of stringencies (Figure \ref{fig:performance_HiC}d).

It should be noted that the IMR90 cell type for which the Hi-C data was available was not part of the cell types from which we predicted the DHS--gene interactions, suggesting that the interactions which could be validated here are not cell-type-specific and that the measured performance is likely an underestimate of the true performance.

\subsubsection{Carbon-Copy 3C (5C) validation}
\label{sec:carbon-copy-3c}

The 5C technique makes use of specific primers to identify chromatin interactions. In \cite{sanyal2012} this technique was used to reveal interactions between TSSs of genes and distal elements in 1\% of the human genome. These sets of interactions, consisting of 2 sets of primers and 3 cell types, are publicly available in the ENCODE database. From this 5C data we constructed a `gold standard' (see Methods for details). The limitation to 1\% of the genome resulted in only a small overlap between the gold standard space and our predictions (Supplementary Table \ref{stab:5c_numbers}). After all filtering, only 1.65\% of our predictions can be validated with this data. Thus a systematic validation using true and false positive rates, similar to the one with the Hi-C data, is not applicable in this case.

We therefore considered each gene that occurred in both the prediction and the validation set separately and counted the number of predicted interactions (\# P) and the number of correctly predicted interactions (\# C), i.e.\ that were also found by the 5C technique. Table \ref{tab:5C_anecdotal} shows these numbers for the genes that occur in all datasets, i.e.\ all intermediate predictions and the gold standard (see Supplementary Table \ref{stab:5C_anecdotal_full} for a full list of genes). After all filtering one third of the predicted interactions was positively validated by the 5C data. Furthermore, it is clear that, since the percentages of correctly predicted interactions increase (i.e.\ 25.30\%, 31.25\% and 33.33\% for respectively the initial, the ISMA filtered and the transcript filtered predictions), the filtering steps indeed improve the quality of the results.

\begin{table}[ht!]
  \centering
    \begin{tabular}{|l|r|r|r|r|r|r|}
    \hline
          & \multicolumn{2}{|c|}{\textbf{initial}} & \multicolumn{2}{|c|}{\textbf{ISMA}} & \multicolumn{2}{|c|}{\textbf{transcript}} \\
    \hline
    \textbf{gene} & \textbf{\# P} & \textbf{\# C} & \textbf{\# P} & \textbf{\# C} & \textbf{\# P} & \textbf{\# C} \\
		\hline
    \textit{CAV2}  & 30    & 2     & 24    & 2     & 20    & 1 \\
    \textit{CTGF}  & 13    & 13    & 12    & 12    & 10    & 10 \\
    \textit{MAP1A} & 4     & 2     & 4     & 2     & 4     & 2 \\
    \textit{MET}   & 28    & 3     & 18    & 3     & 13    & 3 \\
    \textit{MOXD1} & 2     & 1     & 2     & 1     & 2     & 1 \\
    \textit{SELENBP1} & 2     & 0     & 2     & 0     & 1     & 0 \\
    \textit{SERPINB7} & 4     & 0     & 2     & 0     & 1     & 0 \\
		\hline
		\hline
    \textbf{total} & \textbf{83}    & \textbf{21}    & \textbf{64}    & \textbf{20}    & \textbf{51}    & \textbf{17} \\
		\hline
    \end{tabular}
		\caption{\textbf{Validation with 5C data.} Only genes that were present in all datasets, i.e.\ all intermediate results of our calculations and the 5C dataset, are depicted here. \# P represents the number of predictions, \# C the number of `correct' predictions with respect to the 5C data. After the first calculation step 25.30\% of the interactions are predicted correctly with respect to the 5C validation set. After ISMA filtering this increases to 31.25\%, and after transcript filtering to 33.33\%.}
  \label{tab:5C_anecdotal}
\end{table}

\subsection{Cap Analysis of Gene Expression (CAGE) confirms predicted interactions}

CAGE is an experimental technique to identify the promoters and TSS of genes (\citet{shiraki2003}), which has been extensively used within the FANTOM research projects. Recently, in a FANTOM5 research, it has been discovered that, next to small RNA fragments around the TSSs, CAGE also finds small fragments of possible enhancer sites (\citet{andersson2014}). From this data significant enhancer-promoter interactions were predicted based on expression correlation between all pairs of enhancers and promoters within a distance of 500 kbp. This set of predicted interactions was compared with our predictions, as significant overlap between these two sets might assure the validity of both prediction methods. 

\begin{figure}[ht!]
	
  \includegraphics[width=\linewidth]{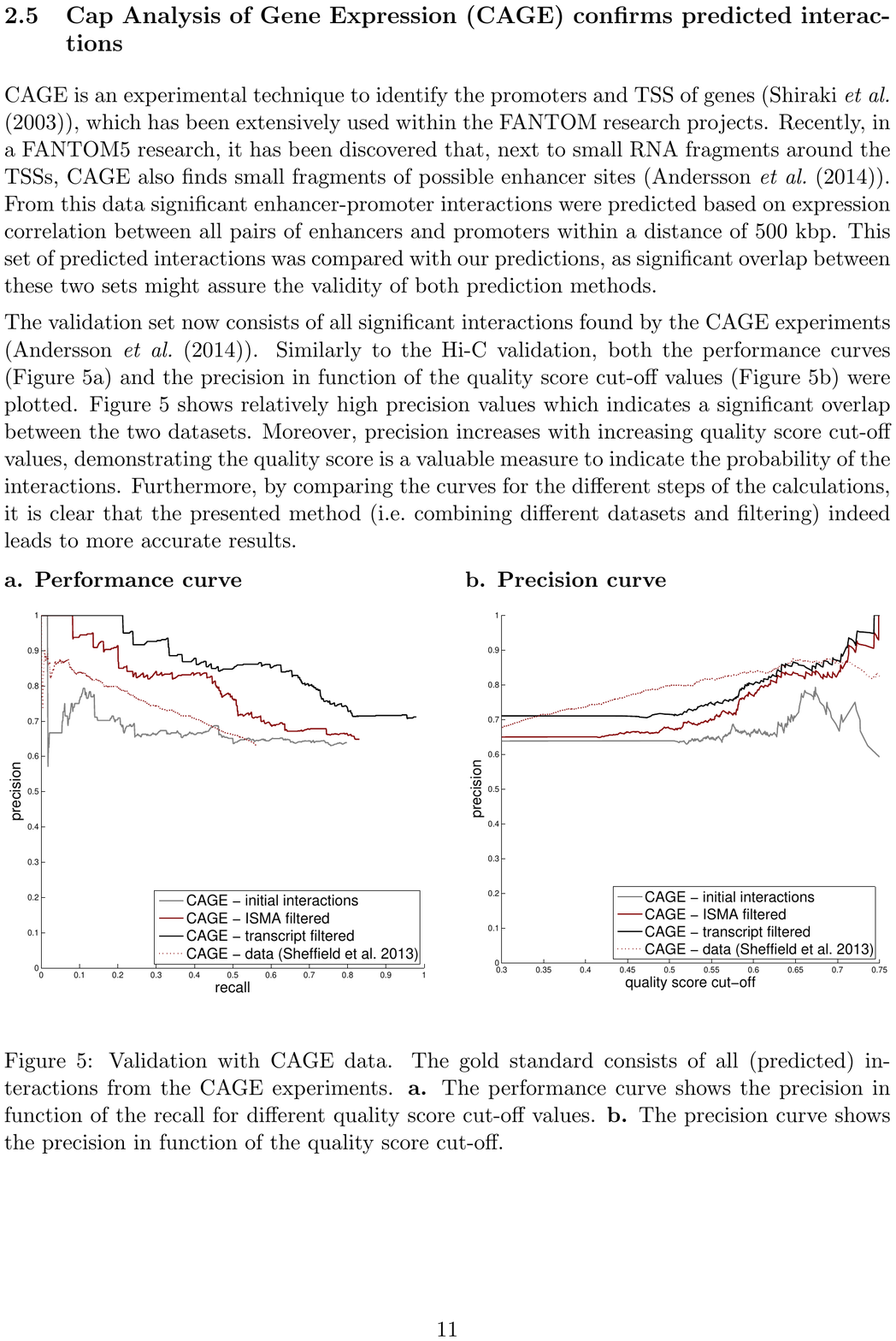}
  
  \caption{\textbf{Validation with CAGE data.} The gold standard consists of all (predicted) interactions from the CAGE experiments. \textbf{a.} The performance curve shows the precision in function of the recall for different quality score cut-off values. \textbf{b.} The precision curve shows the precision in function of the quality score cut-off.}
  \label{fig:CAGE}
\end{figure}

The validation set now consists of all significant interactions found by the CAGE experiments (\citet{andersson2014}). Similarly to the Hi-C validation, both the performance curves (Figure \ref{fig:CAGE}a) and the precision in function of the quality score cut-off values (Figure \ref{fig:CAGE}b) were plotted. Figure \ref{fig:CAGE} shows relatively high precision values which indicates a significant overlap between the two datasets. Moreover, precision increases with increasing quality score cut-off values, demonstrating the quality score is a valuable measure to indicate the probability of the interactions. Furthermore, by comparing the curves for the different steps of the calculations, it is clear that the presented method (i.e.\ combining different datasets and filtering) indeed leads to more accurate predictions.

\subsection{Comparison with known long-range interactions}

\begin{figure}[ht!]
  \centering
  \includegraphics[width=0.8\linewidth]{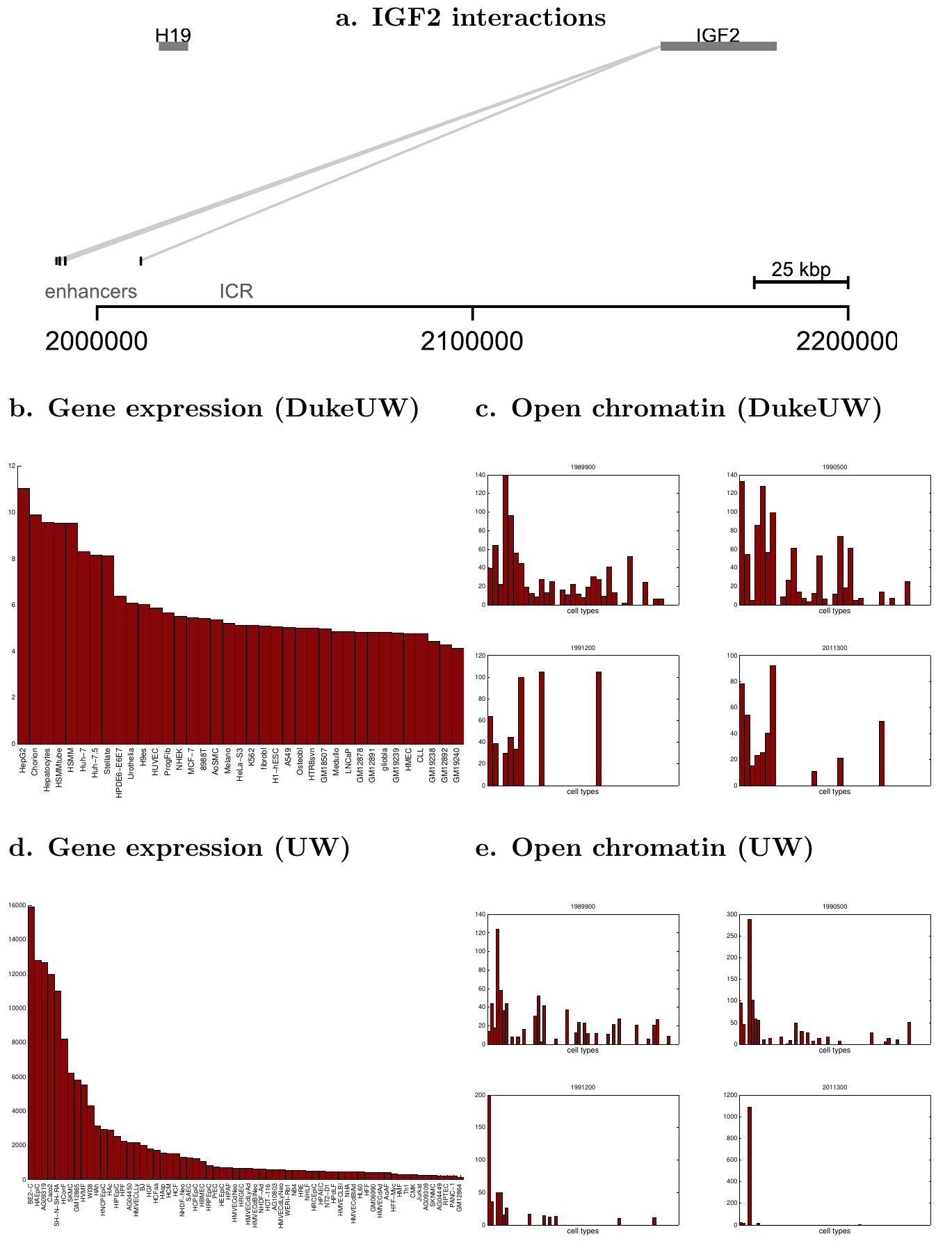}
  \caption{\textbf{Predictions show interaction between the \textit{IGF2} gene and 4 DHSs located upstream of the \textit{H19} promoter.} These interactions are confirmed by the resemblance between the gene expression profiles ((b) and (d) for DukeUW and UW respectively) and the open chromatin profiles of the 4 DHSs ((c) and (e) for DukeUW and UW respectively), i.e.\ higher peaks for the same cell types.}
    \label{fig:IGF2_interactions}
\end{figure}

The well-studied \textit{H19/IGF2} locus has been reported to have an imprinted long-range interaction.  In the original study (\citet{leighton1995}), carried out on mice, it was demonstrated that the \textit{H19} and \textit{IGF2} genes (located on chromosome 7) exhibit parent-of-origin-specific mono-allellic expression. While \textit{H19} is expressed from the maternal chromosome, \textit{IGF2} is expressed from the paternal one. Both genes share enhancer elements and are controlled, besides by the imprinted control region (ICR) situated between the 2 genes, by regions located upstream of the \textit{H19} promoter. This has also been observed in human cell lines (\citet{tabano2010}), in which these genes are located on chromosome 11.

Our results show that there is indeed a long-range interaction between the \textit{IGF2} gene and open chromatin regions situated upstream from the \textit{H19} promoter (Figure \ref{fig:IGF2_interactions}a). This was also reported in \citet{sheffield2013}. Although it is reported that these correlations were primarily driven by liver lineages, this long-range interaction is also observed when comparing multiple cell lines. 

Figures \ref{fig:IGF2_interactions}b-e show the gene expression profiles of the \textit{IGF2} gene for the different cell lines of both datasets, together with the DNase sensitivity profiles of the DHSs in question. These profiles show a fair amount of similarity as the higher peaks are observed for the same cell types.

\subsection{Exploring the data}
To query the predicted interactions, a webservice was developed: http://dhsgen.roslin.ed.ac.uk. By selecting a gene, for which interactions were predicted, from the drop-down list all interacting DHSs can be queried, together with the quality scores of the corresponding interactions. This list of interacting DHSs can be downloaded as a file. In addition, a link to the UCSC Genome Browser is provided in which the interactions are displayed visually.

\section{Discussion}

One of the principal findings of the ENCODE project has been that regulatory elements occupy a much greater portion of the genome than previously anticipated, but understanding how these elements coordinate the precise spatio-temporal regulation of gene expression remains a formidable challenge. A promising approach to link regulatory elements to their candidate target genes uses guilt-by-association: if the `activity' (e.g. DNA accessibility or protein-binding frequency) of a regulatory element and expression level of a gene correlate significantly across multiple experimental conditions or cell types, an interaction between them is inferred. Here we improved on existing approaches in two directions. Firstly, we used a subgraph matching algorithm to identify consistent correlations in multiple datasets of matching DNase-seq and exon array samples. This enabled us to incorporate more samples in our analysis while avoiding the need for complex cross-dataset normalization. Secondly, we also considered gene co-expression interactions in our analysis in order to filter interactions between regulatory elements and target genes that are most likely due to indirect effects, again using a subgraph matching approach. This removed the need to limit our search to the area around genes and allowed us to predict interactions across whole chromosomes.

A critical issue when computationally predicting thousands of interactions is to validate them on a correspondingly large scale. We borrowed validation principles from the network reverse-engineering field and showed that there was a significant overlap between our predictions and genome-wide chromosome capture (Hi-C) data as well as predictions derived from CAGE data. This overlap moreover increased when more stringent thresholds were applied to either the predicted interactions or validation data.

Although the various high-throughput experimental technologies used to generate both the training and validation data each have their own biases and limitations, the extent of overlapping interactions derived from either of them is highly encouraging and suggests that integrating more data types (e.g.  FAIR, ChIP-seq or RNA-seq data) will only lead to more accurate predictions of long-range regulatory interactions. We believe graph-based data integration methods such as the ones introduced here will play a key role in this endeavor.

\section{Methods}

\subsection{Data collection and pre-processing}
Both the DNase-seq and the exon array data was collected from the ENCODE database (\citet{encode2011}). For all used cell types (see Supplementary Table 1) the DNase-seq peak files were downloaded in bigbed-format and translated to the readable bed-format. Similarly, the exon array data was downloaded for both data sources, i.e. DukeUW (collected data from the Crawford lab of Duke University and the Stamatoyannopoulous lab of the University of Washington) and UW (data from the University of Washington). Where possible, summarized data files were used in which the data of the different experiments is collected. All this data made use of the human genome assembly hg19. This resulted in both DNase-seq and DukeUW exon array data for 37 cell types, and DNase-seq and UW exon array data for 66 different cell types.

The DNase-seq data was preprocessed as follows. The whole genome was divided in 100 bp bins, similar to (\citet{degner2012}), and for each bin $B_i$ a single open chromatin level $b_i$ was calculated as a weighted average of all open chromatin peaks located on this bin:

$$b_i = \sum_{j} \frac{D_j \cap B_i}{S} d_j$$

in which $(D_j \cap B_i)$ represents the number of overlapping base pairs between the DNase peak $D_j$ and the bin $B_i$, $S$ is the bin size (in this case 100 bp) and $d_j$ represent the open chromatin level of DNase peak $D_j$.

Subsequently, for each chromosome (except the sex chromosomes) and for each data source, a gene expression matrix and an open chromatin matrix were generated with respectively the gene expression and the open chromatin levels. The gene expression matrix is a $M\times N$ matrix, with $M$ the number of cell types in the data source and $N$ the number of genes/exons in the chromosome. The open chromatin matrix is a $M\times K$ matrix, with $M$ as defined previously and $K$ the number of bins in the chromosome. Supplementary table \ref{stab:number_matrix} shows for each chromosome the dimensions of these matrices.

\subsection{Calculating correlations}
For each chromosome and for each data source, absolute Spearman correlations were calculated between the columns of the open chromatin matrix and the columns of the gene expression matrix. To limit the calculations only the genes/exons for which the expression levels vary sufficiently across the different cell types, i.e.\ genes $g_i$ for which $\text{std}(g_i) > \text{mean}_{n=1..N} (\text{std}(g_n))$, are taken into account, resulting in a $M\times N'$ gene expression matrix. Similarly, only those bins that have open chromatin in at least one of the cell types, are taken into account, resulting in a $M\times K'$ open chromatin matrix. Calculating the correlations results in a $K'\times N'$ correlation matrix (See Supplementary Table \ref{stab:number_matrix} for the values of $N'$ and $K'$).

Subsequently, the interactions with the most significant correlations were selected by applying a false discovery rate (FDR) threshold of 10\%. For this, the absolute Spearman correlations between a random open chromatin matrix (i.e. the open chromatin matrix with the rows permuted randomly) and the gene expression matrix were calculated $n$ times (in our case $n=3$). Then a correlation cut-off value was determined, so that only 10\% of the remaining correlations is due to randomness, i.e.\ applying this cut-off value to the random correlation matrix results in only 10\% of the interactions in comparison to applying this cut-off value to the original correlation matrix. All interactions with a significant correlation, i.e.\ higher than the determined cut-off value, were then kept in a list in which each entry consists of a gene/exon, an open chromatin area and a quality score, represented by the absolute correlation value.

\subsection{ISMA Filtering}
A weighted edge-colored graph was built form the resulting interactions of phase 1 for each dataset, together with location information of the genes/exons of each dataset. All interactions are translated to edges between open chromatin bins and genes/exons, in which each data source is represented by a different edge type (i.e.\ color). Moreover, edges are created between genes and exons that are co-located, i.e.\ alignment links. Weights were assigned to all edges. While for the interaction links this is the correlation value, for the alignment links this is a unit weight, i.e.\ 1.

Subsequently, the Index-based Subgraph Matching Algorithm (ISMA) (\citet{demeyer2013}) was applied in this graph to find all subgraphs representing a significant interaction between a bin and a gene/exon that occurs in both datasets. These subgraphs (see Figure \ref{fig:overview_method}b) are 3-node graphs with 3 different edge types: an interaction between a bin and a gene (dataset 1), an interaction between a bin and an exon (dataset 2) and an alignment between a gene and an exon. When applying the ISMA algorithm a score was calculated for each subgraph by multiplying all edge weights. The square root of this score, which is the geometric mean of the two correlations, is then used as the quality score of the interaction.

\subsection{Clustering and Transcript Filtering}
The motifs (i.e.\ subgraphs) found by the ISMA algorithm are clustered in order to find graph structures similar to the one depicted in figure \ref{fig:overview_method}c. All subgraphs with the same open chromatin region and the same gene are collected together. Each of this clusters represents a single interaction between a DHS and a gene.

Subsequently, this list of clusters is filtered making use of the publicly available A-MEXP-2246 annotation file\footnote{http://www.ebi.ac.uk/arrayexpress/files/A-MEXP-2246/A-MEXP-2246.additional.1.zip} to only keep those clusters that represent known gene transcripts. Only those interactions are retained for which the set of exons in the cluster contains the majority (i.e.\ $>80\%$) of exons of a known gene transcript.

Finally, the quality score of the interaction was calculated as the maximum quality score of all motifs (i.e.\ 3-node subgraphs) participating in the corresponding cluster.

\subsection{Indirect Filtering}
To eliminate indirect interactions (see Figure \ref{fig:indirect_interactions}a), the absolute Spearman correlation was calculated between each pair of genes/exons. Subsequently, a network was constructed with the set of nodes consisting of all DHSs and genes/exons and two types of links: interaction links between DHSs and genes/exons and correlations between genes/exons. The ISMA algorithm was again applied to enumerate all subgraphs with a configuration as depicted in figure \ref{fig:indirect_interactions}b. Subsequently, in each of these subgraphs the link with the lowest weight was identified and if this was an interaction link, the corresponding interaction was removed from the predicted interactions.

Next to this, after all filtering, interactions between DHSs and co-located genes were removed from the predictions.

\subsection{Validation with Hi-C and CAGE data}
Firstly, both the Hi-C and the CAGE validation set were retrieved from their corresponding publications (\citet{jin2013, andersson2014}).

For fair comparison, our set of predicted interactions was limited to only those in the same distance range and both sets were restricted to only those interactions between open chromatin areas and genes which occur in both sets (similar to \cite{stolovitzky2009lessons}). Moreover, the open chromatin areas were translated to the same level of detail. While we opted for 100 bp bins to represent the open chromatin areas, the validation sets utilize different open chromatin regions. The most detailed information is translated to the least detailed. For example, the regions of the Hi-C data contain on average 17,540 bp. This means that each of our DHS bins needed to be translated to the region (of the Hi-C data) in which it is located.

Subsequently, for different quality score cut-off values the set of predicted interactions was compared with the validation set and the precision and recall were calculated as follows:

$$
precision =  \frac{TP}{TP+FP}
$$
$$
recall = \frac{TP}{TP+FN}
$$

with $TP$ the number of true positives, $FP$ the number of false positives and $FN$ the number of false negatives. Plotting the precision in function of the recall for different quality score cut-off values results in the performance curve. Similarly, to investigate the avail of the quality score the precision was plotted in function of the quality score cut-off.

\subsection{Validation with 5C data}
\label{ssec:validation_5c}
The 5C data, as published in (\cite{sanyal2012}), was collected from the ENCODE database. This dataset consists of 5C data for 3 cell types (GM12878, K562 and HeLa-S3) and 2 sets of primers. For each set of primers, all interactions of the different cell types were collected and those interactions that are present in at least 2 cell types were added to the gold standard.

Subsequently, for each gene we counted the number of predicted interactions and the number of `correctly' predicted interactions. Hereby, we limited the set of predictions to only those interactions of which both the gene and the DHS are present in the gold standard.

\section{Data Access}
All the data used in this research is publicly available. Both the exon array and the DNase-seq data were collected from the ENCODE database. Both the Hi-C and the CAGE validation data was gathered from the corresponding publications. The 5C validation set was collected from the ENCODE database. The predicted interactions can be queried at http://dhsgen.roslin.ed.ac.uk.

\section{Acknowledgements}
We thank David Hume for providing us with early access to the CAGE data and results. We thank Andy Law for the help with the webservice. This research was supported by Roslin Institute Strategic Grant funding from the BBSRC.

% \section{Figures}
% \listoffigures

% \section{Tables}
% \listoftables 

% \bibliographystyle{msb}
% \bibliography{bibsysbiol,references}

\newpage

\appendix

\setcounter{equation}{0}
\renewcommand{\theequation}{S\arabic{equation}}
\setcounter{figure}{0}
\renewcommand{\thefigure}{S\arabic{figure}}
\setcounter{table}{0}
\renewcommand{\thetable}{S\arabic{table}}

\section{Supplementary Figures}

\begin{figure}[ht!]
    \centering
			%\begin{subfigure}{\textwidth}
					%\includegraphics[width=0.9\textwidth]{DHS_ifo_TSS_full1.eps}
					%\caption{initial interactions}
					%\label{sfig:DHS_1}
			%\end{subfigure}
			%\begin{subfigure}{\textwidth}
					%\includegraphics[width=0.9\textwidth]{DHS_ifo_TSS_full2.eps}
					%\caption{ISMA filtered}
					%\label{sfig:DHS_2}
			%\end{subfigure}
			%\begin{subfigure}{\textwidth}
					%\includegraphics[width=0.9\textwidth]{DHS_ifo_TSS_full3.eps}
					%\caption{transcript filtered}
					%\label{sfig:DHS_3}
			%\end{subfigure}
			
    \includegraphics[width=0.78\linewidth]{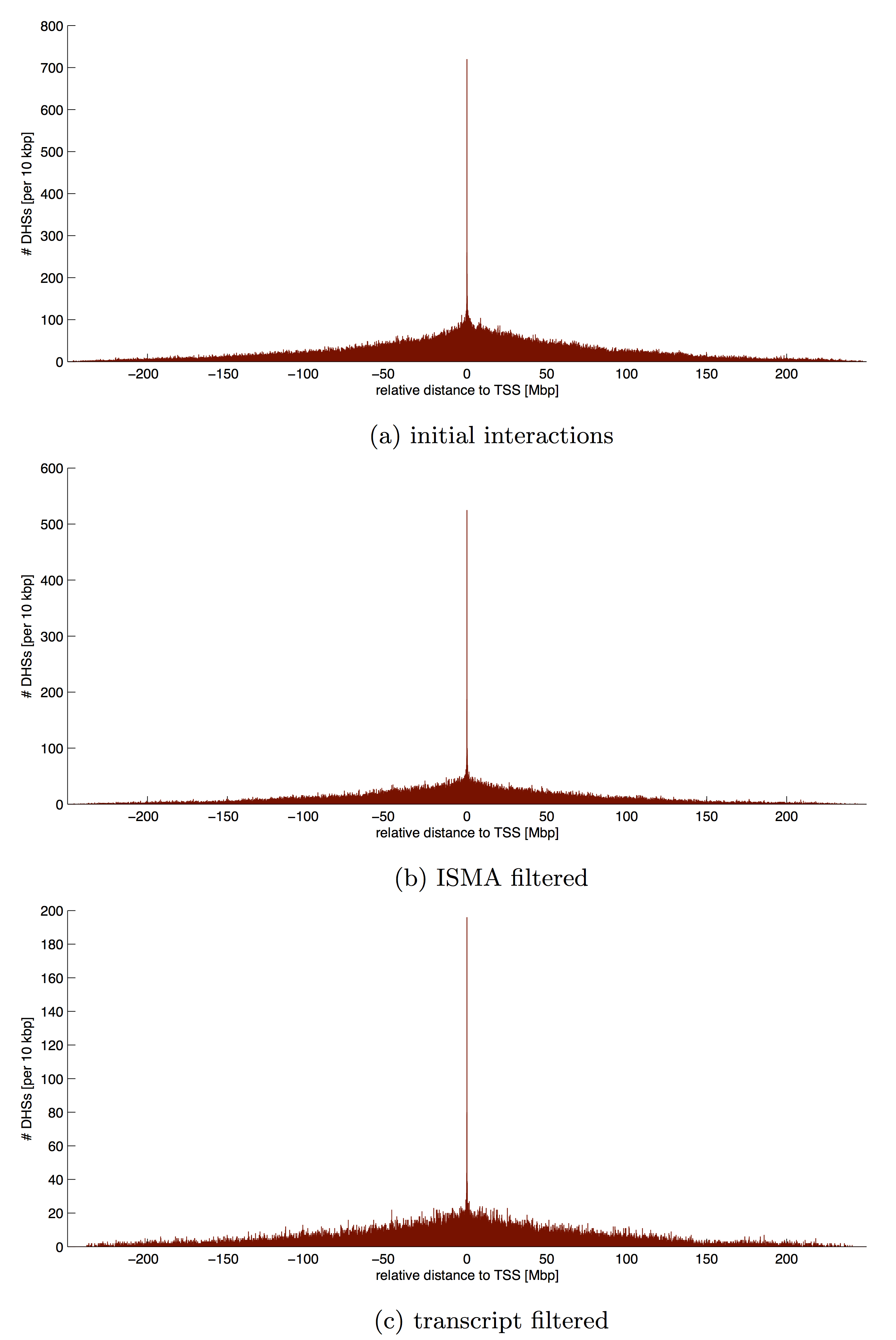}
			
    \caption{Number of open chromatin areas in function of the relative distance to the TSS of the genes they are interacting with. (a) initial interactions (DukeUW). (b) after ISMA filtering. (c) after transcript filtering. All indirect interactions were eliminated from this data. While the highest concentrations are situated around the TSSs, still long-range interactions ($>$100Mb) are found, and this in all steps of the calculations.}
    \label{sfig:DHS_ifo_TSS}
\end{figure}

\begin{figure}[ht!]
    \centering
			%\begin{subfigure}{\textwidth}
					%\includegraphics[width=0.9\textwidth]{boxplot_distance_QS_full1.eps}
					%\caption{initial interactions}
					%\label{sfig:QS_box_1}
			%\end{subfigure}
			%\begin{subfigure}{\textwidth}
					%\includegraphics[width=0.9\textwidth]{boxplot_distance_QS_full2.eps}
					%\caption{ISMA filtered}
					%\label{sfig:QS_box_2}
			%\end{subfigure}
			%\begin{subfigure}{\textwidth}
					%\includegraphics[width=0.9\textwidth]{boxplot_distance_QS_full3.eps}
					%\caption{transcript filtered}
					%\label{sfig:QS_box_3}
			%\end{subfigure}
			
    \includegraphics[width=0.8\linewidth]{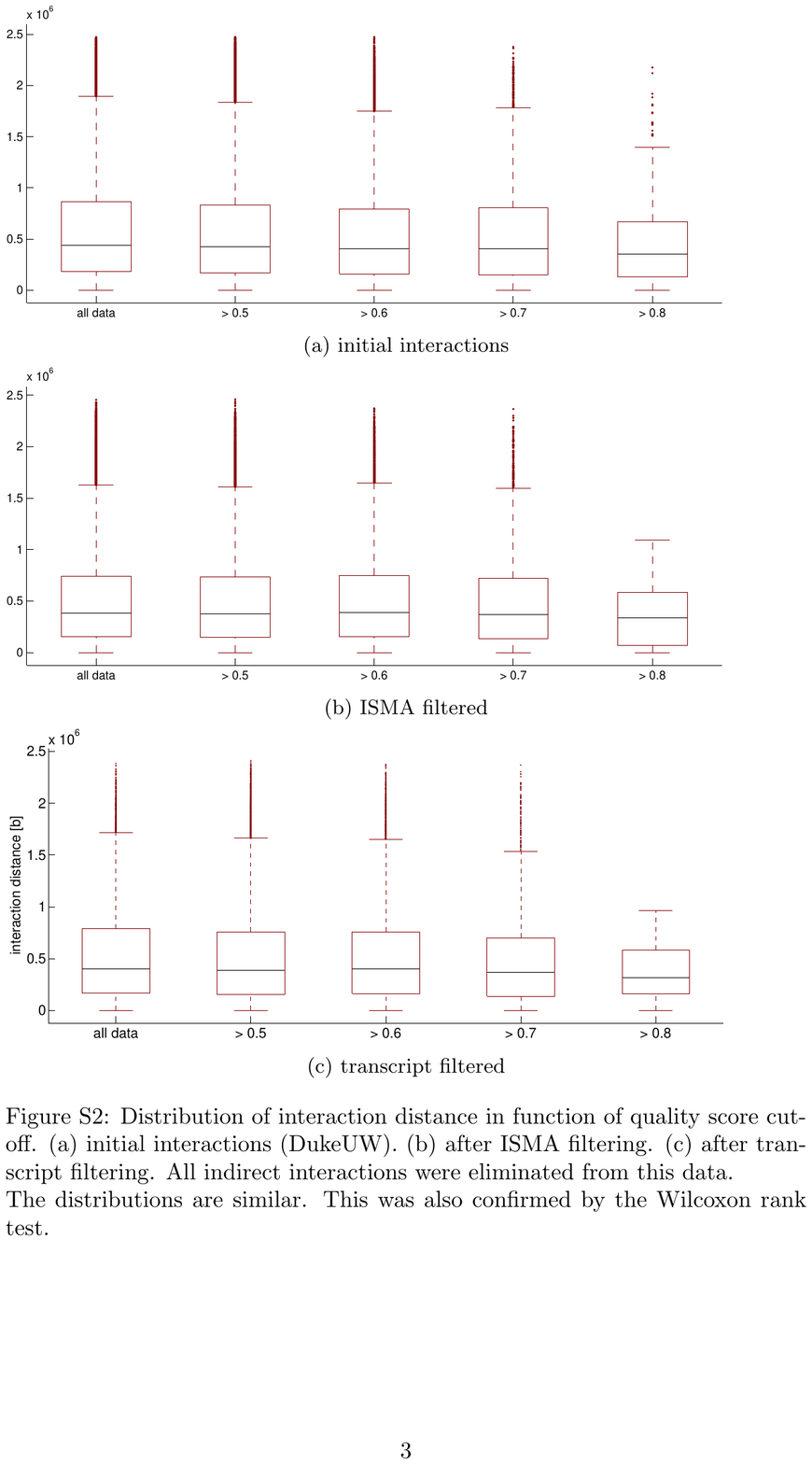}
			
    \caption{Distribution of interaction distance in function of quality score cut-off. (a) initial interactions (DukeUW). (b) after ISMA filtering. (c) after transcript filtering. All indirect interactions were eliminated from this data. The distributions are similar. This was also confirmed by the Wilcoxon rank test.}
    \label{sfig:QS_box}
\end{figure}

\begin{figure}[ht!]
    \centering
			%\begin{subfigure}{\textwidth}
					%\includegraphics[width=0.9\textwidth]{QS_heatmap_1.eps}
					%\caption{initial interactions}
					%\label{sfig:QS_1}
			%\end{subfigure}
			%\begin{subfigure}{\textwidth}
					%\includegraphics[width=0.9\textwidth]{QS_heatmap_2.eps}
					%\caption{ISMA filtered}
					%\label{sfig:QS_2}
			%\end{subfigure}
			%\begin{subfigure}{\textwidth}
					%\includegraphics[width=0.9\textwidth]{QS_heatmap_3.eps}
					%\caption{transcript filtered}
					%\label{sfig:QS_3}
			%\end{subfigure}
			
    \includegraphics[width=0.8\linewidth]{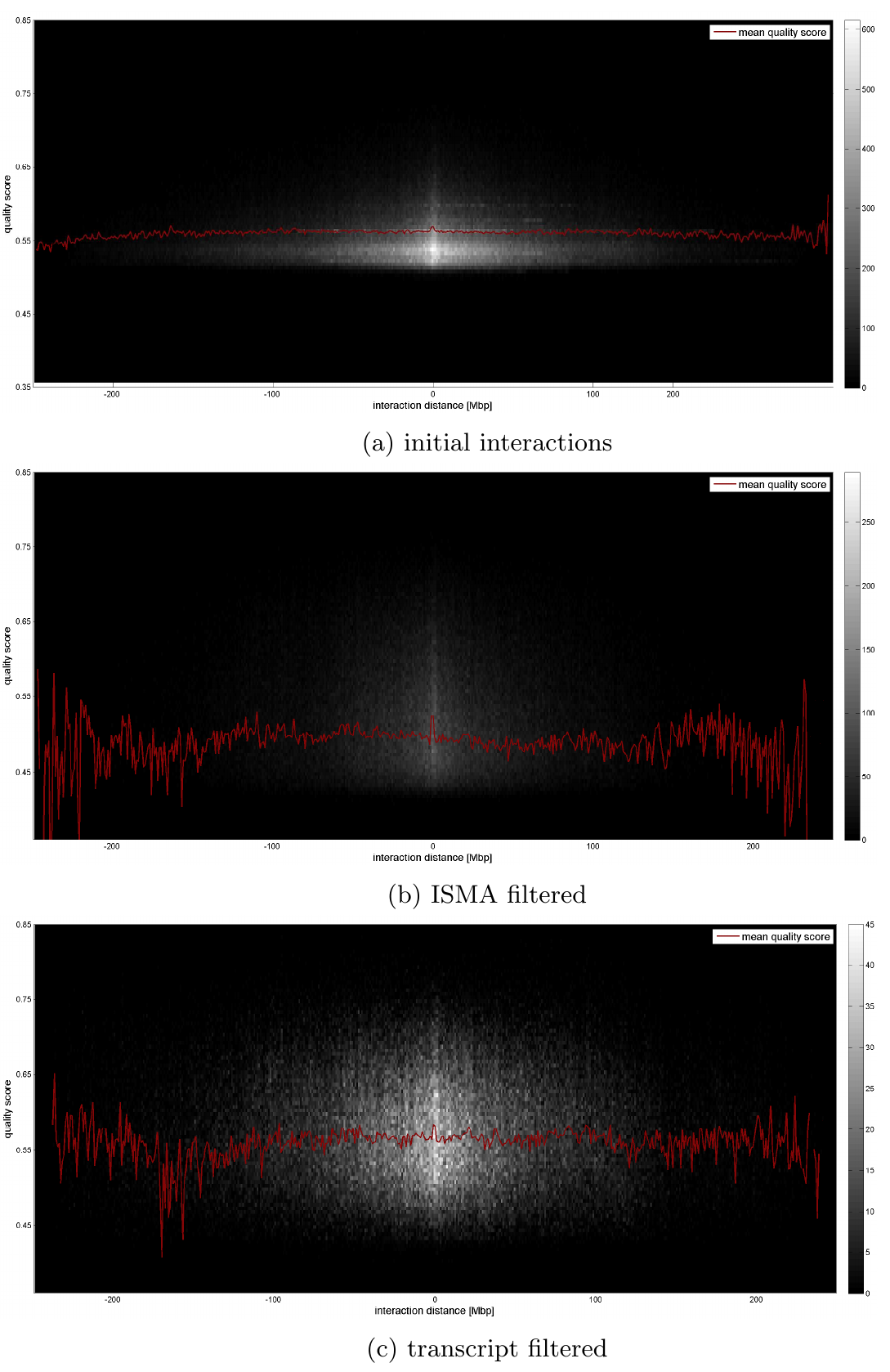}
			
    \caption{Heat map representing the number of interactions in function of both the quality score and the interaction distance. Moreover, the mean quality score is depicted in function of the interaction distance. (a) initial interactions (DukeUW). (b) after ISMA filtering. (c) after transcript filtering. All indirect interactions were eliminated from this data.\\ The mean quality score does not decrease with an increasing interaction distance.}
    \label{sfig:QS_ifo_TSS}
\end{figure}

\begin{figure}[ht!]
    \centering
		\includegraphics[width=\linewidth]{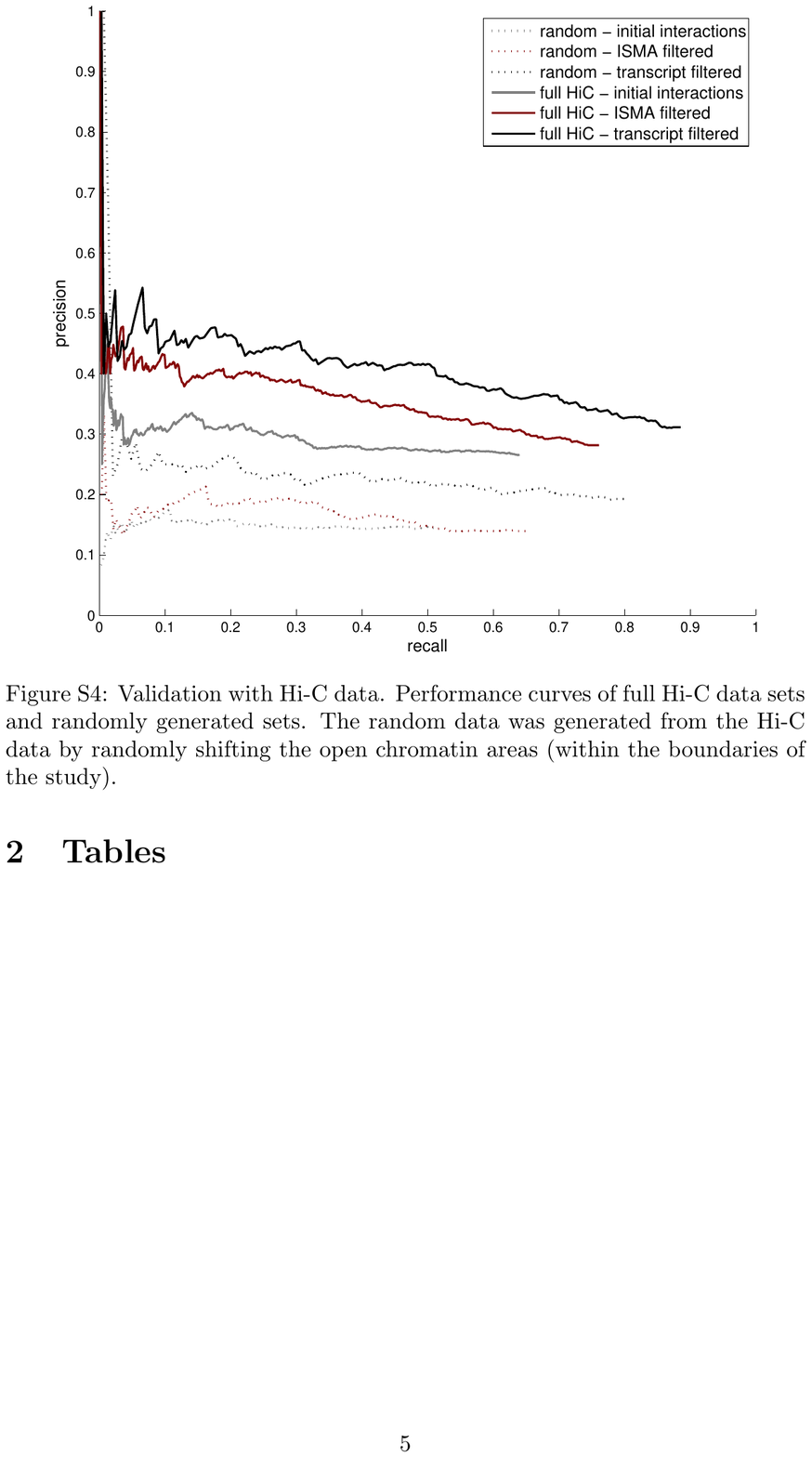}
    \caption{Validation with Hi-C data. Performance curves of full Hi-C data sets and randomly generated sets. The random data was generated from the Hi-C data by randomly shifting the open chromatin areas (within the boundaries of the study).}
    \label{sfig:performance_HiC_random}
\end{figure}

\FloatBarrier

\newpage
\section{Supplementary Tables}

\begin{table} [h!]
\begin{center}
\rotatebox{90}{
\begin{tabular}[h!]{|l|c|}
\hline
\textbf{dataset}	& \textbf{cell types}\\
\hline
\hline
\textbf{DukeUW}	& 
\begin{tabular}{cccccc}
GM12878 & H1-hESC & K562 & A549 & HeLa-S3 & HepG2\\ 
HUVEC & MCF-7 & 8988T & AoSMC & Chorion & CLL\\ 
fibrobl & gliobla & GM12891 & GM12892 & GM18507 & GM19238\\
GM19239 & GM19240 & H9es & Hepatocytes & HMEC & HPDE6-E6E7\\
HSMM & HSMMtube & HTR8svn & Huh-7 & Huh-7.5 & LNCaP\\
Medullo & Melano & NHEK & Osteobl & ProgFib & Stellate\\
Urothelia & & & & & \\
\end{tabular}\\
\hline
\textbf{UW} & 
\begin{tabular}{cccccc}
AG04449 & AG04450 & AG09309 & AG09319 & AG10803 & AoAF\\
BE2\_C & BJ & Caco2 & CMK & GM06990 & GM12864\\
GM12865 & HAc & HAEpiC & HAh & Hasp & HBMEC\\ 
HCFaa & HCF & HCM & HConF & HCPEpiC & HCT-116\\
HEEpiC & HFF & HFF-Myc & HGF & HIPEpiC & HL60\\
HMF & HMVECdAd & HMVECdBlAd & HMVECdBlNeo & HMVECdLyAd & HMVECdLyNeo\\
HMVECdNeo & HMVECLBl & HMVECLLy & HNCPEpiC & HPAEC & HPAF\\
HPdLF & HPF & HRCEpiC & HRE & HRGEC & HRPEpiC\\
HVMF & Jurkat & NB4 & NHA & NHDF-Ad & NHDF-Neo\\
NHLF & NT2-D1 & PANC-1 & PrEC & RPTEC & SAEC\\
SKMC & SKNMC & SH-N-SH\_RA & Th1 & WERI\_Rb1 & WI38\\
\end{tabular}\\
\hline
\end{tabular}
}
\end{center}
\caption{Cell types used in this study. For each cell type both DNase-seq and exon array data was available in the ENCODE database. While the \textbf{DukeUW} dataset has \textbf{37} cell types, the \textbf{UW} dataset has \textbf{66}.}
\label{stab:cell_types}
\end{table}

\begin{table}
\begin{center}
\rotatebox{90}{
\begin{tabular}[h!]{|l|r|r|r|r|r|}
\hline
\textbf{chr} &	\textbf{initial (DukeUW)} &	\textbf{initial (UW)} &	\textbf{after ISMA filtering}	& \textbf{after clustering} &	\textbf{after transcript filtering}\\
\hline
\textbf{1} &	1 421 108	& 151 119 224	& 3 357 575	& 371 894	& 84 737\\
\textbf{2}	& 685 762	& 119 479 384	& 3 241 181	& 245 717	& 84 954\\
\textbf{3}	& 442 461	& 72 603 469	& 1 597 656	& 149 125	& 37 472\\
\textbf{4}	& 253 948	& 46 103 987	& 855 367	& 87 340	& 15 868\\
\textbf{5}	& 378 351	& 57 004 407	& 1 744 283	& 155 633	& 32 488\\
\textbf{6}	& 579 172	& 65 068 838	& 1 546 938	& 146 525	& 32 413\\
\textbf{7}	& 242 378	& 46 901 260	& 971 176	& 79 527	& 26 407\\
\textbf{8}	& 234 535	& 34 569 307	& 885 225	& 86 014	& 24 830\\
\textbf{9}	& 297 096	& 31 769 158	& 873 564	& 92 810	& 23 400\\
\textbf{10}	& 301 967	& 35 605 487	& 648 361	& 92 745	& 19 652\\
\textbf{11}	& 399 316	& 37 295 059	& 1 002 805	& 130 492	& 25 240\\
\textbf{12}	& 455 633	& 46 877 221	& 1 271 164	& 138 754	& 31 419\\
\textbf{13}	& 75 244	& 12 469 439	& 164 971	& 22 841	& 4 564\\
\textbf{14}	& 280 416	& 18 733 392	& 655 965	& 63 938	& 15 029\\
\textbf{15}	& 204 513	& 23 443 870	& 787 587	& 70 929	& 13 630\\
\textbf{16}	& 57 737	& 20 296 943	& 165 902	& 21 729	& 4 846\\
\textbf{17}	& 395 977	& 31 105 653	& 707 914	& 99 358	& 11 369\\
\textbf{18}	& 51 268	& 8 255 982	& 149 129	& 18 326	& 4 451\\
\textbf{19}	& 189 316	& 13 259 671	& 254 455	& 44 949	& 3 997\\
\textbf{20}	& 105 156	& 10 165 970	& 187 969	& 29 376	& 4 623\\
\textbf{21}	& 12 782	& 3 379 461	& 68 758	& 4 824	& 1 281\\
\textbf{22}	& 64 912	& 5 199 664	& 96 657	& 14 826 &	3 046\\
\hline
\textbf{total}	& \textbf{7 129 048}	& \textbf{890 706 846} &	\textbf{21 234 602}	& \textbf{2 167 672}	& \textbf{505 716}\\
\hline
\end{tabular}
}
\end{center}
\caption{Number of `interactions' per chromosome after each step in the calculations.}
\label{stab:number_interactions}
\end{table}

\begin{table}
\begin{center}
\rotatebox{90}{
\begin{tabular}[h!]{|l|r|r|r|r|r|r|}
\hline
\textbf{chr} &	\textbf{initial} &	\textbf{\parbox{0.20\textwidth}{initial \\+ indirect}} &	\textbf{\parbox{0.20\textwidth}{ISMA \\+ cluster}}	& \textbf{\parbox{0.20\textwidth}{ISMA \\+ cluster \\+ indirect}} &	\textbf{transcript} & \textbf{\parbox{0.20\textwidth}{transcript \\+ indirect}}\\
\hline
\textbf{1} &	1 421 108	& 129 542	&  371 894 & 34 798	& 84 737 & 16 007\\
\textbf{2}	& 685 762	& 77 737	&  245 717 & 20 924	& 84 954 & 10 205\\
\textbf{3}	& 442 461	& 47 361	&  149 125 & 13 739	& 37 472 & 6 014\\
\textbf{4}	& 253 948	& 39 853	&  87 340	& 16 364 & 15 868 & 6 063\\
\textbf{5}	& 378 351	& 52 332	&  155 633 & 24 253	& 32 488 & 8 933\\
\textbf{6}	& 579 172	& 69 677	&  146 525	& 19 634 & 32 413 & 6 743\\
\textbf{7}	& 242 378	& 29 306	&  79 527	& 16 873 & 26 407 & 8 307\\
\textbf{8}	& 234 535	& 36 203	&  86 014	& 17 338 & 24 830 & 8 246\\
\textbf{9}	& 297 096	& 35 643	&  92 810	& 18 520 & 23 400 & 6 924\\
\textbf{10}	& 301 967	& 45 508	& 92 745	& 21 275 & 19 652 & 7 788\\
\textbf{11}	& 399 316	& 45 808	& 130 492	& 19 802 & 25 240 & 7 920\\
\textbf{12}	& 455 633	& 48 744	& 138 754	& 25 231 & 31 419 & 11 476\\
\textbf{13}	& 75 244	& 14 528	& 22 841	& 7 604 & 4 564 & 2 995\\
\textbf{14}	& 280 416	& 28 339	& 63 938	& 12 807 & 15 029 & 5 011\\
\textbf{15}	& 204 513	& 23 058	& 70 929	& 15 304 & 13 630 & 6 401\\
\textbf{16}	& 57 737	& 9 667	& 21 729	& 5 493 & 4 846 & 2 345\\
\textbf{17}	& 395 977	& 46 117	& 99 358	& 18 531 & 11 369 & 5 693\\
\textbf{18}	& 51 268	& 16 880	& 18 326	& 8 043 & 4 451 & 2 716\\
\textbf{19}	& 189 316	& 26 752	& 44 949	& 9 362 & 3 997 & 2 315\\
\textbf{20}	& 105 156	& 19 814	& 29 376	& 9 013 & 4 623 & 2 613\\
\textbf{21}	& 12 782	& 2 294	& 4 824	& 2 052 & 1 281 & 827\\
\textbf{22}	& 64 912	& 9 264	& 14 826 &	4 712 & 3 046 & 1 866\\
\hline
\textbf{total}	& \textbf{7 129 048}	& \textbf{854 427} &	\textbf{2 167 672}	& \textbf{341 672} & \textbf{505 716} & \textbf{137 408}\\
\hline
\end{tabular}
}
\end{center}
\caption{Number of interactions per chromosome before and after indirect filtering. The initial interactions are those of the DukeUW dataset. After filtering out the indirect interactions, 11.99\% of the initial interaction remain, 15.76\% of the interactions after ISMA filtering, and 27.17\% after transcript filtering.}
\label{stab:number_indirect_interactions}
\end{table}

\begin{table}
\begin{center}
\begin{tabular}[h!]{|l|r|r|r|r|}
\hline
\textbf{\parbox{0.16\textwidth}{interaction \\distance}} &	\textbf{\parbox{0.17\textwidth}{interactions \\(DukeUW)}} &	\textbf{\parbox{0.17\textwidth}{interactions \\(UW)}} &\textbf{\parbox{0.18\textwidth}{interactions \\after ISMA \\filtering}}	& \textbf{\parbox{0.22\textwidth}{interactions \\after transcript \\filtering}}\\
\hline
$>$100kb & 99.40\% & 99.60\% & 99.13\% & 99.07\%\\
$>$1Mb & 97.36\% & 97.84\% & 96.86\% & 96.82\%\\
$>$50Mb & 47.26\% & 47.88\% & 47.58\% & 48.22\%\\
$>$100Mb & 21.27\% & 20.49\% & 21.21\% & 20.90\%\\
$>$200Mb & 1.24\% & 1.26\% & 1.25\% & 1.06\%\\
\hline
\end{tabular}
\end{center}
\caption{Percentages of `long-range' interactions for different cut-off distances. A relative high percentage (i.e.\ ~20\%) of the interactions are long-range interactions ($>$100Mb)}
\label{stab:long_range}
\end{table}

\begin{table}
\begin{center}
\rotatebox{90}{
\begin{tabular}[h!]{|l||r|r||r|r||r|r|r|}
\hline
\textbf{chr} &	\textbf{N (DukeUW)} &	\textbf{N' (DukeUW)} &	\textbf{N (UW)}	& \textbf{N' (UW)} &	\textbf{K} & \textbf{K' (DukeUW)} & \textbf{K' (UW)}\\
\hline
\textbf{1} &	3 808	& 1 499	& 123 871 & 26 462	& 2 492 507 & 476 111 & 399 815\\
\textbf{2}	& 2 672	& 1 066	&  113 087 & 22 195	& 2 431 994 & 418 787 & 374 972\\
\textbf{3}	& 2 203	& 857	& 83 766 & 17 003	& 1 980 225 & 315 318 & 295 917\\
\textbf{4}	& 1 749	& 689	& 66 650	& 13 307 & 1 911 543 & 224 180 & 238 344\\
\textbf{5}	& 1 939	& 778	& 72 358  & 14 301	& 1 809 153 & 279 877 & 252 860\\
\textbf{6}	& 2 322	& 924	& 75 791	& 15 176 & 1 711 151 & 285 489 & 265 142\\
\textbf{7}	& 1 973	& 787	& 75 944 & 15 291 & 1 591 387 & 275 561 & 234 964\\
\textbf{8}	& 1 634	& 679	& 60 775 & 11 824 & 1 463 641 & 236 856 & 215 215\\
\textbf{9}	& 1 479	& 600	& 56 179	& 12 025 & 1 412 135 & 220 126 & 185 432\\
\textbf{10}	& 1 582	& 620 & 67 506 & 13 672 & 1 355 348 & 264 850 & 211 668\\
\textbf{11}	& 2 411	& 958 & 68 321 & 14 046 & 1 350 066 & 261 722 & 217 536\\
\textbf{12}	& 2 088	& 818 & 64 701 & 13 150 & 1 338 519 & 241 391 & 215 934\\
\textbf{13}	& 976	& 405 & 37 238	& 7 158 & 1 151 699 & 125 190 & 131 338\\
\textbf{14}	& 1 526	& 596 & 44 450 & 9 097 & 1 073 496 & 164 553 & 134 581\\
\textbf{15}	& 1 175	& 451 & 48 684 & 10 265 & 1 025 314 & 176 321 & 138 797\\
\textbf{16}	& 1 027	& 392 & 50 321	& 10 617 & 903 548 & 187 695 & 132 554\\
\textbf{17}	& 1 595	& 595 & 58 2887 & 13 084 & 811 953 & 228 856 & 166 414\\
\textbf{18}	& 402	& 157 & 31 090	& 6 017 & 780 773 & 129 494 & 102 607\\
\textbf{19}	& 1 665	& 641 & 48 338	& 11 091 & 591 290 & 169 061 & 110 051\\
\textbf{20}	& 1 072	& 433 & 33 415 & 7 030 & 630 256 & 164 596 & 119 048\\
\textbf{21}	& 532	& 226 & 16 811	& 3 339 & 481 299 & 68 423 & 58 144\\
\textbf{22}	& 881	& 333 & 25 009 &	5 630 & 513 046 & 118 402 & 76 190\\
\hline
\end{tabular}
}
\end{center}
\caption{Sizes of gene expression matrices (MxN) and open chromatin matrices (MxK) before and after selecting the most important genes (N') and the significant DHSs (K'). M represents the number of cell types (M=37 for DukeUW, M=66 for UW). N represent the number of genes, K the number of DHSs.}
\label{stab:number_matrix}
\end{table}

\begin{table}[ht!]
  \centering
    \begin{tabular}{|l|r|r|r|}
    \hline
    \textbf{predictions} & \textbf{total number} & \textbf{overlap with} & \textbf{overlap with} \\
				 &  & \textbf{gold standard space} & \textbf{gold standard }\\
    \hline
    \textbf{initial} & 13 871 & 190 & 57\\
		\textbf{ISMA filtered} & 7 790 & 111 & 34\\
		\textbf{transcript filtered} & 3 099 & 51 & 17\\
		\hline
    \end{tabular}
		\caption{The number of interactions within the range of the 5C data (i.e.\ $<$1Mb), the number of interactions that could be validated with this data (i.e.\ present in the gold standard space), and hte number of `correct' predictions with respect to the 5C data.}
  \label{stab:5c_numbers}
\end{table}

\begin{table}[ht!]
  \centering
    \begin{tabular}{|l|r|r|r|r|r|r|}
    \hline
          & \multicolumn{2}{|c|}{\textbf{initial}} & \multicolumn{2}{|c|}{\textbf{ISMA}} & \multicolumn{2}{|c|}{\textbf{transcript}} \\
    \hline
    \textbf{gene} & \# P & \# 'C' & \# P & \# 'C' & \# P & \# 'C' \\
		\hline
    \textit{AP000304.12} & 1     & 0     & -     & -     & -     & - \\
    \textit{AP000936.3} & 1     & 1     & -     & -     & -     & - \\
    \textit{AP003774.5} & 1     & 1     & -     & -     & -     & - \\
    \textit{APOC3} & 1     & 0     & -     & -     & -     & - \\
    \textit{CAV2}  & 30    & 2     & 24    & 2     & 20    & 1 \\
    \textit{CRAT}  & 4     & 0     & -     & -     & -     & - \\
    \textit{CTGF } & 13    & 13    & 12    & 12    & 10    & 10 \\
    \textit{EIF6}  & 1     & 1     & -     & -     & -     & - \\
    \textit{FGF1}  & 2     & 2     & -     & -     & -     & - \\
    \textit{HBZ}   & 6     & 2     & -     & -     & -     & - \\
    \textit{HOXA2} & 4     & 2     & 3     & 1     & -     & - \\
    \textit{HOXA9} & 4     & 3     & 1     & 0     & -     & - \\
    \textit{IRF1 } & 4     & 1     & -     & -     & -     & - \\
    \textit{LILRB1} & 5     & 2     & -     & -     & -     & - \\
    \textit{MAP1A} & 4     & 2     & 4     & 2     & 4     & 2 \\
    \textit{MET}   & 28    & 3     & 18    & 3     & 13    & 3 \\
    \textit{MOXD1 }& 2     & 1     & 2     & 1     & 2     & 1 \\
    \textit{OR52B6} & 1     & 0     & -     & -     & -     & - \\
    \textit{P4HA2} & 26    & 3     & 20    & 2     & -     & - \\
    \textit{PGC}   & 1     & 0     & -     & -     & -     & - \\
    \textit{POLR3K} & 1     & 1     & -     & -     & -     & - \\
    \textit{RAD50} & 1     & 1     & -     & -     & -     & - \\
    \textit{RP11-298J23.6 }& 2     & 2     & -     & -     & -     & - \\
    \textit{SELENBP1 }& 2     & 0     & 2     & 0     & 1     & 0 \\
    \textit{SERPINB13} & 1     & 0     & -     & -     & -     & - \\
    \textit{SERPINB2} & 2     & 2     & 2     & 2     & -     & - \\
    \textit{SERPINB7} & 4     & 0     & 2     & 0     & 1     & 0 \\
    \textit{SLC22A4} & 26    & 7     & 19    & 7     & -     & - \\
    \textit{ST7}   & 5     & 0     & -     & -     & -     & - \\
    \textit{TUFT1} & 5     & 5     & 2     & 2     & -     & - \\
		\hline
		\hline
    \textbf{total} & \textbf{190}    & \textbf{57}    & \textbf{111 }   & \textbf{34}    & \textbf{51}    & \textbf{17} \\
		\hline
          &       & 0.3000 &       & 0.3063 &       & 0.3333 \\
		\hline
    \end{tabular}
		\caption{For all genes, present in both the predictions and the gold standard space, the number of predicted interactions (\#P) and the number of `correct' predictions (\#C) with respect to the 5C data.}
  \label{stab:5C_anecdotal_full}
\end{table}

\end{document}